\documentclass[aps,prl,twocolumn,showpacs,showkeys,superscriptaddress]{revtex4}
\usepackage[dvips]{graphicx}
\usepackage{amsmath}
\usepackage{amssymb}
\usepackage[mathcal]{eucal}
\usepackage{mathrsfs}
\usepackage{subfigure}

\newcommand{\jj}{$J_1-J_2$}
\newcommand{\vect}[1]{\boldsymbol{#1}}
\newcommand{\jone}{$J_1$}
\newcommand{\jtwo}{$J_2$}
\newcommand{\neel}{N\'{e}el}
\newcommand{\pb}{$\mathrm{Pb}_2\mathrm{VO(PO}_4\mathrm{)}_2$}
\newcommand{\sr}{$\mathrm{SrZn}\mathrm{VO(PO}_4\mathrm{)}_2$}

\newcommand{\li}{$\mathrm{Li}_2\mathrm{VOSiO}_4$}

\newcommand{\tn}{$T_\mathrm{N}$}
\newcommand{\uumlaut}{\"{u}}
\newcommand{\xt}{$\chi\mathrm{(T)}$}
\newcommand{\tmax}{$T_\mathrm{Max}$}
\newcommand{\thcw}{$\theta_{\mathrm{CW}}$}

\begin{document}


\title{Spin correlations and exchange in square lattice frustrated ferromagnets}


\author{M. Skoulatos}
\email{markos.skoulatos@helmholtz-berlin.de}
\affiliation{Helmholtz Centre Berlin for Materials and Energy, Glienicker Str. 100, 14109 Berlin, Germany}

\author{J.P. Goff}
\affiliation{Department of Physics, Royal Holloway, University of London, Egham, Surrey TW20 0EX, United Kingdom}

\author{C. Geibel}
\affiliation{Max-Planck Institute for Chemical Physics of Solids, N\"{o}thnitzer Str. 40, Dresden 01187, Germany}

\author{E.E. Kaul}
\affiliation{Max-Planck Institute for Chemical Physics of Solids, N\"{o}thnitzer Str. 40, Dresden 01187, Germany}

\author{R. Nath}
\affiliation{Max-Planck Institute for Chemical Physics of Solids, N\"{o}thnitzer Str. 40, Dresden 01187, Germany}

\author{N.~Shannon}
\affiliation{H.H. Wills Physics Laboratory, University of Bristol, Bristol BS8 1TL, United Kingdom}

\author{B. Schmidt}
\affiliation{Max-Planck Institute for Chemical Physics of Solids, N\"{o}thnitzer Str. 40, Dresden 01187, Germany}

\author{A.P. Murani}
\affiliation{Institut Laue-Langevin, 156X, 38042 Grenoble Cedex, France}

\author{P.P. Deen}
\affiliation{Institut Laue-Langevin, 156X, 38042 Grenoble Cedex, France}

\author{M. Enderle}
\affiliation{Institut Laue-Langevin, 156X, 38042 Grenoble Cedex, France}

\author{A.R. Wildes}
\affiliation{Institut Laue-Langevin, 156X, 38042 Grenoble Cedex, France}



\date{\today}
\begin{abstract}

The \jone-\jtwo~model on a square lattice exhibits a rich variety of 
different forms of magnetic order that depend sensitively on the ratio 
of exchange constants \jtwo/\jone. We use bulk magnetometry and polarized 
neutron scattering to determine \jone~and \jtwo~unambiguously for two 
materials in a new family of vanadium phosphates, \pb~and \sr, and we 
find that they have ferromagnetic \jone. The ordered moment in the 
collinear antiferromagnetic ground state is reduced, and the diffuse 
magnetic scattering is enhanced, as the predicted bond-nematic region of 
the phase diagram is approached.
\end{abstract}

\pacs{75.50.−y; 75.40.Cx; 75.40.Gb; 75.30.Et; 75.25.+z}
\keywords{\jj~model; Diffuse scattering; Collinear antiferromagnet; Ferromagnetic exchange}

\maketitle




%

The square-lattice $S=1/2$ Heisenberg antiferromagnet with nearest-neighbour 
(NN) exchange constant \jone~and next-nearest-neighbour (NNN) exchange 
\jtwo, has long served as a paradigm for the two-dimensional frustrated 
magnetism (for a review see Ref. \cite{misguich_ws04}). 
It is described by the following Hamiltonian:
\begin{equation}
   \label{equ:j1j2ham}
\mathscr{H} = J_1\sum_{<i,j>_1}\vect{S}_{i} \cdot \vect{S}_{j}+J_2\sum_{<i,k>_2}\vect{S}_{i} \cdot \vect{S}_{k}
 \end{equation}
\noindent where $<i,j>_1$ and $<i,k>_2$ refer to pairs of NN and NNN respectively.
The so-called \jj~model first came to prominence due to its relevance to the 
high-temperature superconducting cuprates, and the recent discovery of 
high-\textit{T}c superconductivity in 
pnictides~\cite{kamihara_jacs08,chen_nat08} has increased interest in this 
model, since it may play a key role for the magnetism on the square 
Fe sublattice~\cite{si_prl08}.

\begin{figure}[b]
   \includegraphics[scale=0.4,angle=0]{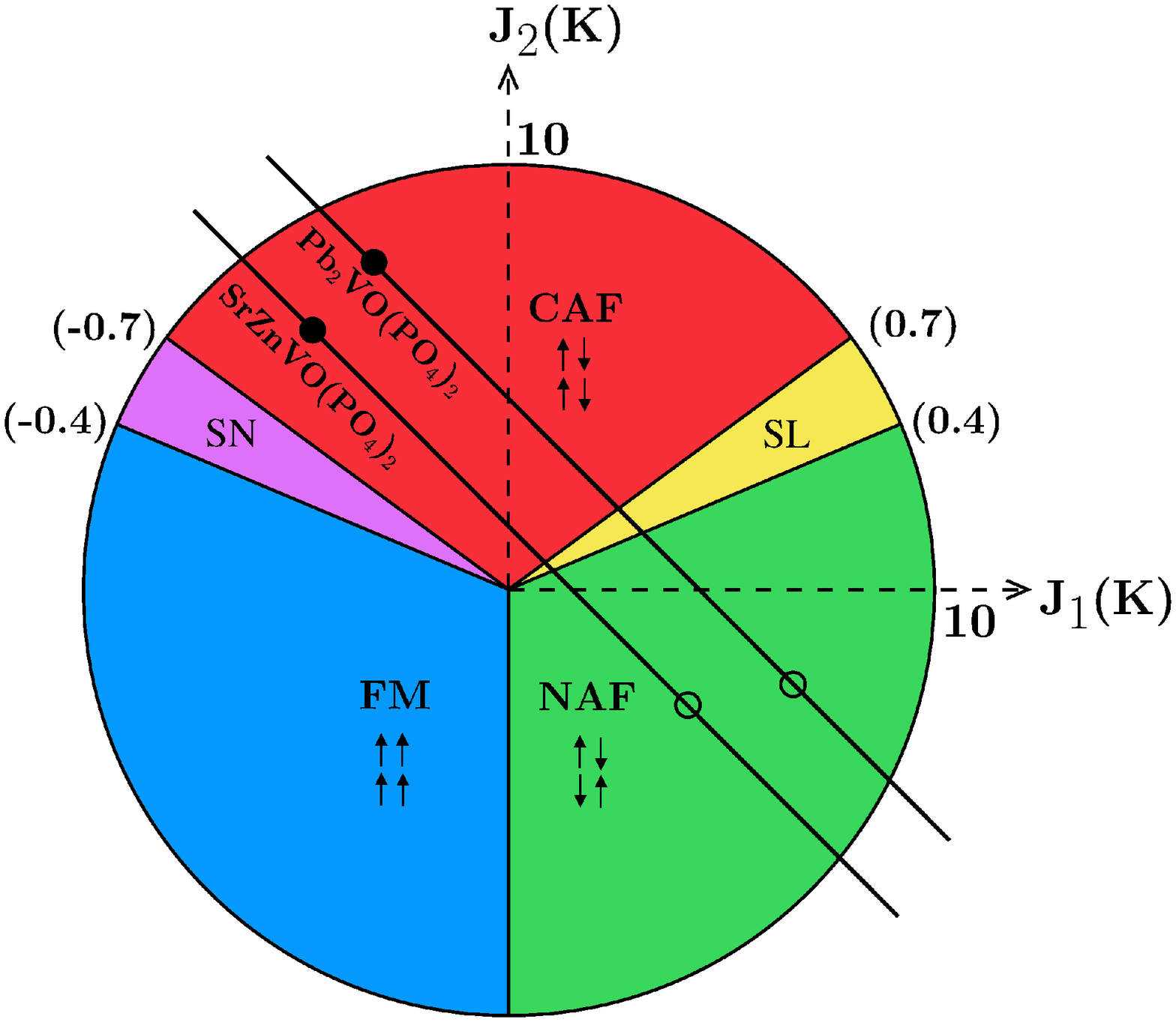}
   \caption{Calculated phase diagram for the general \jj~model, 
   locating \pb~and \sr.
   The values of $\alpha $ in parentheses are the
   phase boundaries determined by exact diagonalization 
   \cite{schulz_jpif96,shannon_prl06}. 
   At these values, zero-point fluctuations destroy the relevant order parameter.
   The yellow area corresponds to the valence bond solid (VBS) regime, the purple 
   zone to the bond-nematic (d-BN) region \cite{shannon_prl06,shannon_epjb04,thalmeier_prb08}.
   The straight lines are obtained from the 
   Curie-Weiss temperatures, and the circles are obtained from the 
   maxima in susceptibility, as described in the text.
   The filled circles are the
   solutions selected using polarized neutrons.\label{fig:phase_diag}}
\end{figure}

The frustrated square lattice model is characterized by the frustration ratio
$\alpha=J_2/J_1$, and the energy scale is given by $J_c = \sqrt{J^2_1 + J^2_2}$. 
When the exchange constants are antiferromagnetic (AF) and $\alpha \approx 0.5$
the ground state is believed to be a valence bond solid (VBS) in which spins 
form tightly bound singlets on nearest-nieghbour bonds~\cite{schulz_jpif96}.  
On the other hand, recent calculations for ferromagnetic (FM) \jone~and AF \jtwo~predict 
$d$-wave bond-nematic order for $\alpha \approx -0.5$~\cite{shannon_epjb04,shannon_prl06}. 
The calculated phase diagram is shown schematically in Fig.~\ref{fig:phase_diag}.

There are now several experimental realizations of spin-1/2 magnets on a square-lattice 
where both \jone and \jtwo~interactions play an important role.   
The first examples, Li$_2$VOXO$_4$ (X = Si, Ge), are collinear antiferromagnets (CAF) with 
AF exchange $J_2 \gg J_1$~\cite{millet_mrb98,melzi_prl00,melzi_prb01,carretta_prl02,carretta_prb02,rosner_prl02,rosner_prb03,bombardi_prl04}, 
whereas VOMoO$_4$ \cite{bombardi_prb05} and PbVO$_3$ 
are \neel~antiferromagnets (NAF) with AF exchange $J_2 \ll J_1$~\cite{tsirlin_prb08,oka_ic08}.
In contrast, the new family of vanadium phosphates AA'VO(PO$_4$)$_2$ (A,A' = Pb, Zn, Sr, Ba, Cd) 
\cite{kaul_jmmm04,shpanchenko_acsc06,skoulatos_jmmm07,nath_prb08} and 
(CuX)LaNb$_2$O$_7$ (X = Cl, Br) \cite{kageyama_jpsj05,oba_jpsj06} have 
been argued to be frustrated ferromagnets with FM \jone~and AF \jtwo.

In order to explore this phase diagram, accurate
control of $\alpha$ is required. Until now 
it has proved difficult to determine the exchange constants accurately using 
bulk thermodynamic properties, and this is reflected by the wide variation 
reported for $\alpha$ for individual compounds in the literature. 
The determination of $\alpha$ from thermodynamic quantities alone is 
undermined by the fact that $J_1 + J_2$ and $|J_1 + J_2|$, but not $J_1-J_2$ are well determined.

Here we present combined bulk magnetization and polarized neutron scattering 
studies of \pb~and \sr. The determination of the magnetic structures in the 
ground state, and the comparison of the diffuse neutron scattering intensity 
in the paramagnetic phase with high-temperature series expansions (HTSE) 
provides an unambiguous determination of the exchange constants \jone~and 
\jtwo. We are also able to explore the consequences for the spin correlations 
of the approach to a quantum disordered region of the phase diagram.

Large powder samples were synthesized at the Max-Planck Institute for Chemical Physics of Solids in Dresden
by the solid state reaction method, in order to avoid multi-phase composition for these compounds. 
The quasielastic neutron experiments were performed on the D7 diffuse scattering spectrometer at the
Institut Laue-Langevin (ILL) in Grenoble, with a fixed incident wavelength of $\lambda=3.1~$\AA~corresponding to an energy window of 8.5 meV.
XYZ polarization analysis was employed to separate magnetic signal from the coherent structural and incoherent scattering \cite{scharpf_pss93}.
Polarized inelastic neutron scattering measurements were performed using the IN20 triple-axis spectrometer of the ILL, in order to confirm the energy scale of the system. Scans of energy transfer were performed at various reciprocal lattice points, $Q$, and temperatures, $T$, and in all cases the excitations were found to be within $\pm4~\mathrm{meV}$.
Thus the measurements on D7 integrate over the full spectral energy range to yield the structure factor, $S(Q)$.
The bulk magnetisations of \pb~and \sr~were measured using a SQUID magnetometer
at the University of Liverpool.
\begin{figure}[b]
  \begin{center}
    \hspace{-3mm}
    \vspace{-5mm}
    \subfigure[]{\label{fig:pbstruct-a}\includegraphics[scale=0.14]{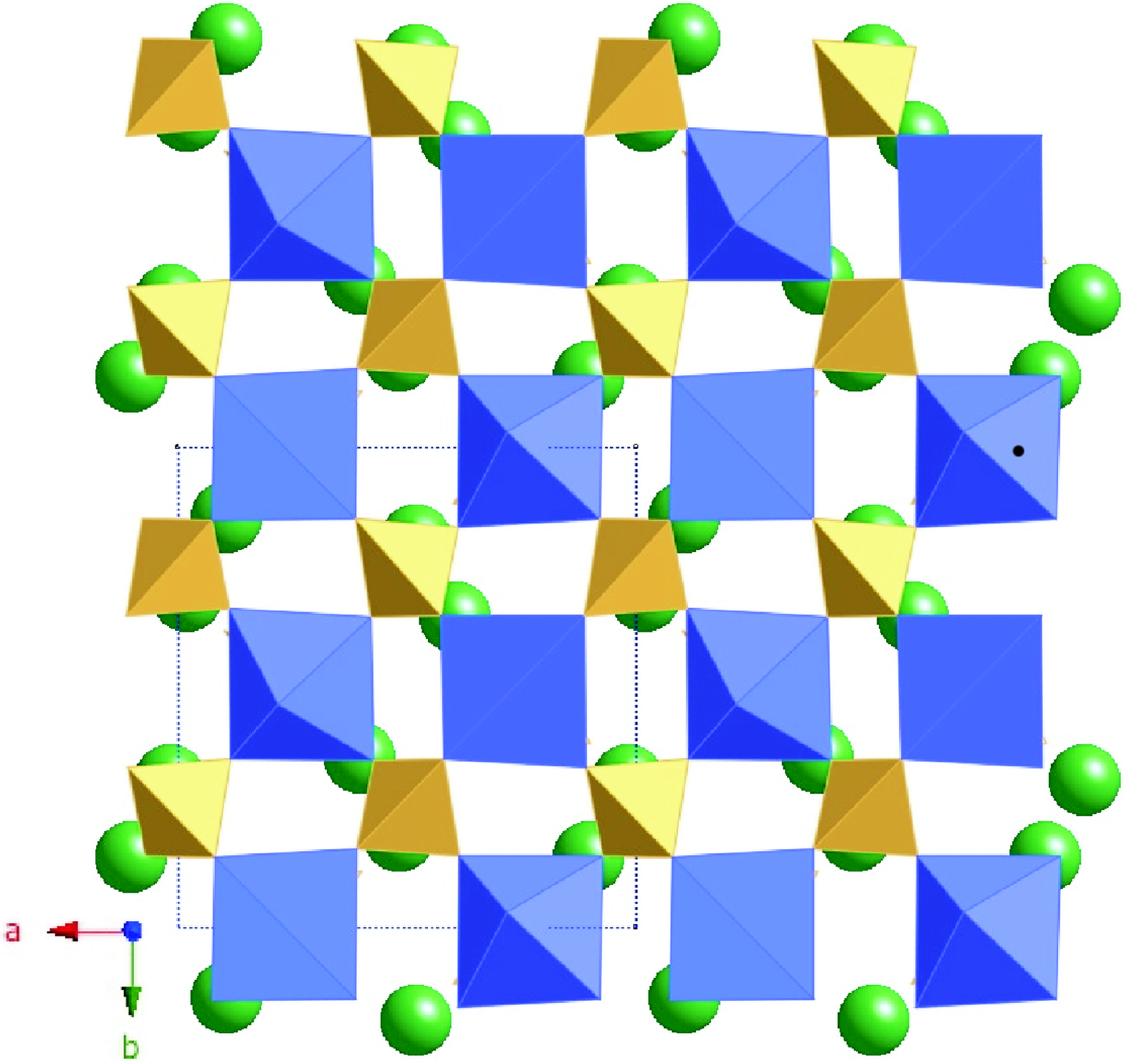}}~~~~~
    \subfigure[]{\label{fig:pbstruct-b}\includegraphics[scale=0.14]{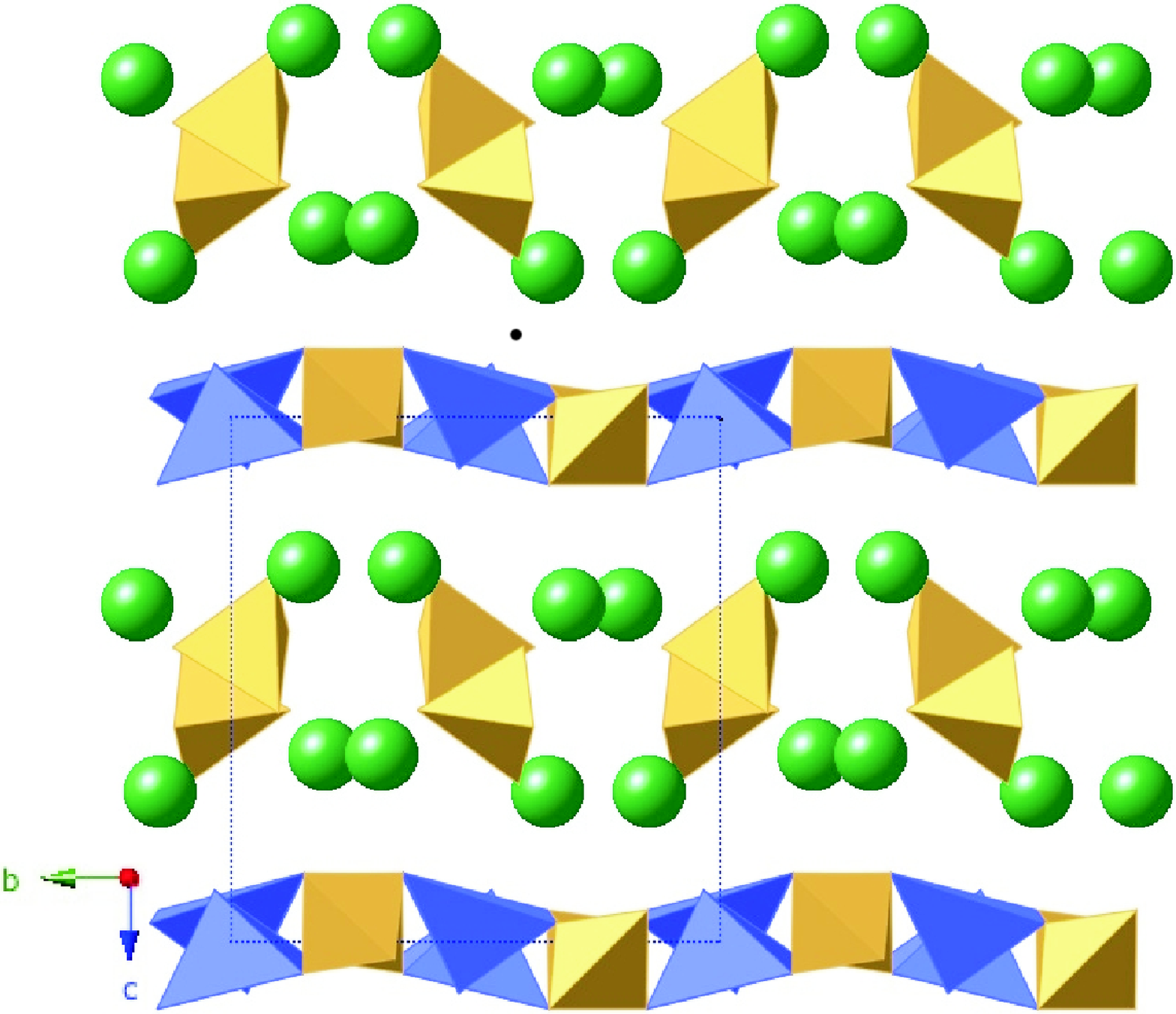}}
  \end{center}
  \caption{The structure of \pb~contains corrugated layers formed by VO$_5$ square pyramids (blue), oriented in
    alternating directions in a chessboard fashion. The pyramids are connected by tetrahedral
    PO$_4$ groups (yellow). The layers are separated by the Pb atoms (green) and isolated PO$_4$ tetrahedra.
    Projections shown are for a) \textit{ab} and b) \textit{bc} planes.\label{fig:pbstruct}}
\end{figure}

\pb~crystallizes in the  monoclinic system with spacegroup \textit{P}$2_1/$\textit{a} and
lattice parameters $a=8.747(4)$~\AA, $b=9.016(5)$~\AA, $c=9.863(9)$~\AA~and
$\beta=100.96(4)^{\circ}$~\cite{kaul_thesis,shpanchenko_acsc06}.
Two projections of the \pb~structure are shown in Fig.~\ref{fig:pbstruct}.
The structure contains corrugated layers formed by VO$_5$ square pyramids oriented in
alternating directions in a chessboard fashion. The pyramids are connected by tetrahedral
PO$_4$ groups.
The layers are buckled along the \textit{b-axis}.
The square bases of the similarly oriented pyramids are located approximately at the same
level.
The magnetic [VOPO$_4$] layers are well separated ($\sim$10\AA ) by non-magnetic Pb atoms
and isolated PO$_4$ tetrahedra, indicating a quasi-2D character for the system.
\sr~shares many characteristics with its relative \pb.
Their structural relationship is the replacement of the two Pb$^{2+}$
cations by Sr$^{2+}$ and Zn$^{2+}$.
Again, it forms a two-dimensional square lattice of $\mathrm{V}^{4+}$ ions
with competing interactions leading to frustration.
The profound structural similarities with \pb~attracted our interest since small
changes introduced by the cationic substitution can tune the magnetic exchanges
\jone~and \jtwo~and thus the frustration ratio.
\sr~crystallises
in the orthorhombic system with spacegroup \textit{Pbca} and
lattice parameters $a=9.0660$~\AA, $b=9.0117$~\AA~and $c=17.5130$~\AA~
\cite{kaul_thesis,meyer_b}.
The main difference between the two compounds' space groups arises from the way the
$\mathrm{VOPO}_{4}$ layers are stacked along the \textit{c-axis}.
\begin{figure}[h]
   \includegraphics[scale=0.62,angle=0]{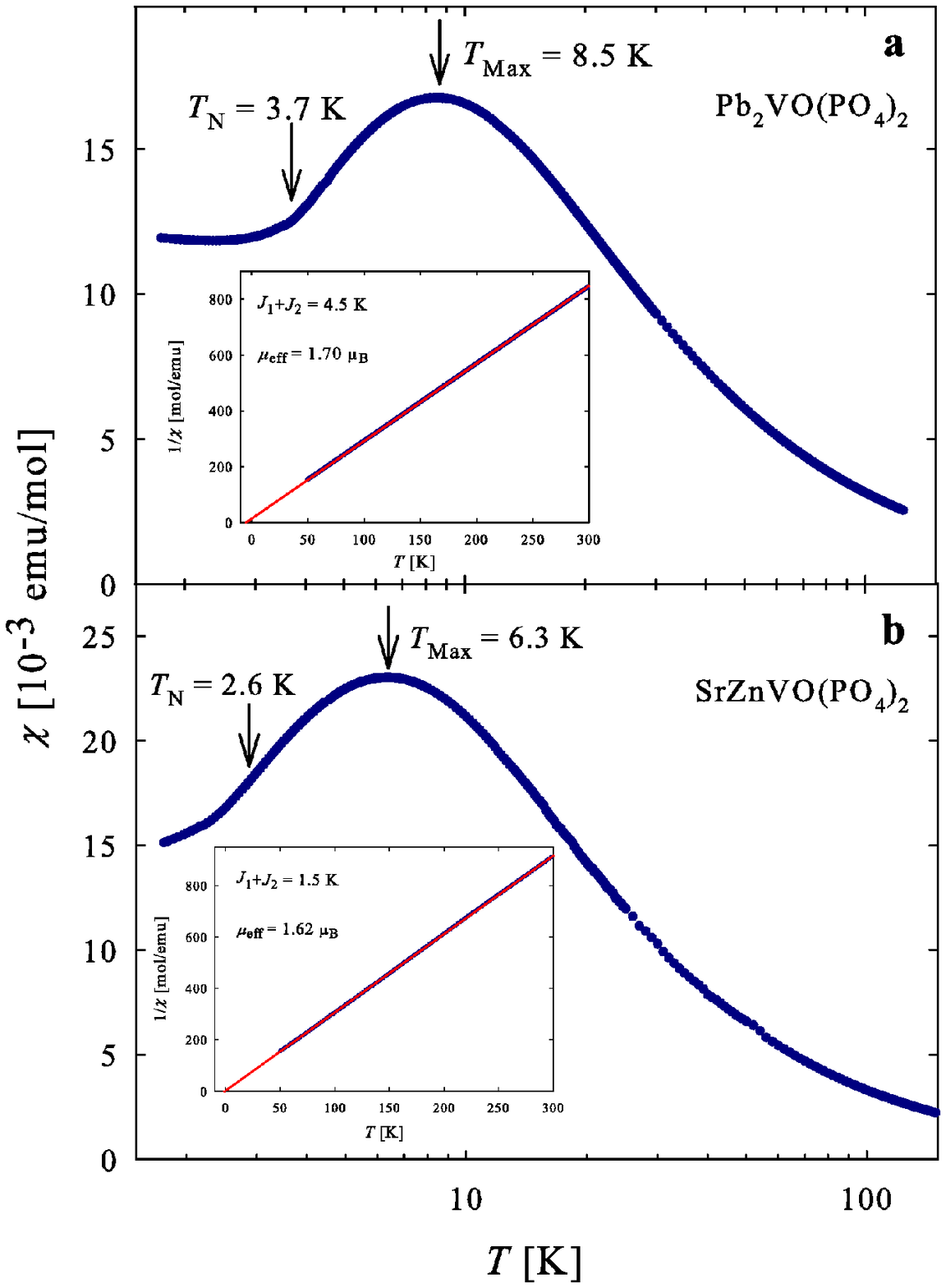}
   \caption{Magnetic response of polycrystalline (a) \pb~and (b) \sr~as measured at 1000 Oe.
   The broad peaks of \xt~are characteristic for low dimensional spin systems.
   The magnetic phase transitions at \tn~manifest as a change of slope in \xt, while
   \tmax~indicate the energy scale for these systems.
   Insets show inverse susceptibilities for $T>50~\mathrm{K}$, together with
   the Curie-Weiss law fits (solid lines), giving \thcw~= (4.5$\pm1$) K and \thcw~= (1.5$\pm1$) K
   for \pb~and \sr~respectively.\label{fig:squid}}
\end{figure}

The polycrystalline samples were measured under an applied field of 1000 Oe, giving the results of Fig.~\ref{fig:squid}.
For T$>$50 K, \xt~is characterised by a paramagnetic Curie-Weiss (CW) behaviour.
Below that, \xt~passes through a broad maximum that is characteristic of a low-dimensional magnet, peaked at \tmax~$\sim$ 8.5 K and 6.3 K for \pb~and \sr~respectively.
The good quality of the samples is evident by the absence of any trace of impurity
tails at the lowest attainable temperatures, which is typical of paramagnetic foreign phases.
Kinks in \xt~indicate the onset of magnetic ordering at the \neel~temperatures
\tn~$\sim$ 3.7 K and 2.6 K for \pb~and \sr~respectively.
The fits of the Curie-Weiss law to the high-temperature data (T$>$50 K) are shown as insets in Fig.~\ref{fig:squid}. The effective moments $\mu_{eff}=(1.72 \pm0.02)$ are consistent with the predicted value for $S=1/2$, $\mu_{eff}=1.73$.
The extrapolated intercepts with the temperature axis give $\theta_{\mathrm{CW}}=J_1+J_2 = (4.5\pm1$) K and (1.5$\pm1$) K for \pb~and \sr~respectively.
The values of \thcw~are rather small compared to the positions of the maxima in the
susceptibilities \tmax, which allow the overall energy scales of
the systems $J_c$ to be estimated.
This suggests that \jone~and \jtwo~ have opposite signs, providing evidence
of mixed FM and AF exchange couplings~\cite{shannon_epjb04}.
The conclusions from the susceptibility measurements for the two compounds are summarized
in the phase diagram of Fig.~\ref{fig:phase_diag}.
The observed value of \thcw~defines a straight line on this plot,
while the relation between \tmax~and $J_c$, which has been calculated
in Ref.~\cite{shannon_epjb04}, enables us to get a pair of solutions on this line.
Thus for both compounds, there is one possible solution in the CAF and another
in the NAF region of the phase diagram.

\begin{figure}[h]
   \includegraphics[scale=0.57,angle=0]{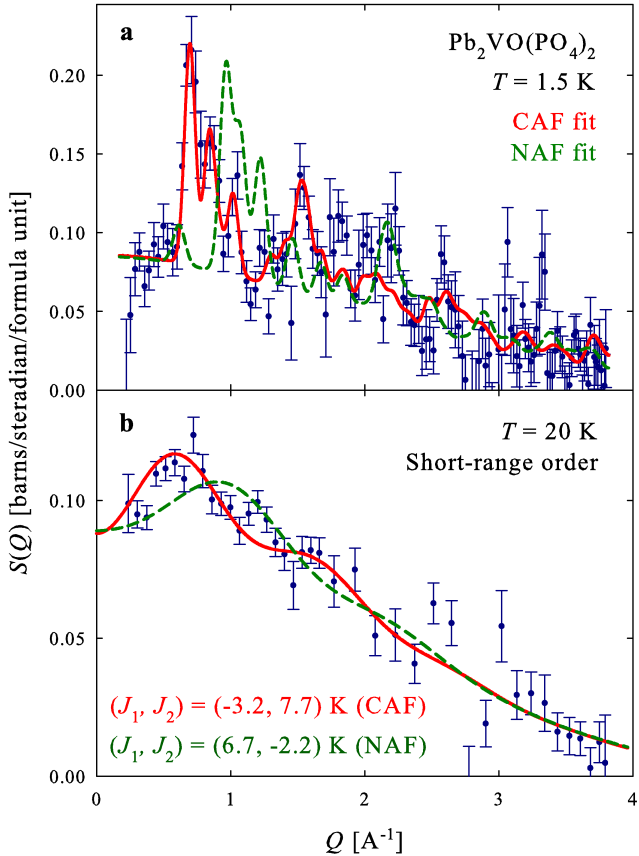}
   \caption{(a) Magnetic ground state diffraction pattern of \pb~measured at 1.5 K.
   The ordered component of the ions' moments gives rise to sharp magnetic Bragg reflections,
   while the disordered component gives a diffuse background.
   Fits are a CAF model with $\vect{Q}=(\pi,0)$ (solid line) as well as a \neel~model
   (dashed line, for comparison only).
   Part (b) shows short-range spin correlations in the paramagnetic phase of \pb.
   The solid curve is a first-order HTSE calculation for the CAF solution with FM $J_1=(-3.2\pm1)~\mathrm{K}$ and AF $J_2=(7.7\pm1)~\mathrm{K}$, and the dashed is the NAF solution.
The data clearly favor the CAF solution.\label{fig:d7}}
\end{figure}

The neutron scattering experiment complemented these bulk property results, yielding a unique solution in each case.
All of the scattering in Fig.~\ref{fig:d7} is magnetic, and it has been placed on an absolute scale by reference to a standard vanadium sample.
At temperatures below the \neel~point, the magnetic scattering
consists of sharp magnetic Bragg reflections on top of a diffuse magnetic background.
Data for powdered samples of \pb, \sr, as well as a control sample of \li~were collected at 1.5 K and were all found to have a CAF structure.
This immediately selects only one of the two solutions consistent with the bulk magnetization data.
Fig.~\ref{fig:d7}a~shows the magnetic ground state diffraction pattern of \pb,
modelled by both a CAF structure with an ordering wavevector of $\vect{Q}=(\pi,0)$ (solid line)
and a \neel~structure for comparison (dashed line that fails to reproduce the data).
The ordered moments in the CAF structures are (0.42$\pm0.04 \mu_B$) for
\sr, (0.50$\pm0.04 \mu_B$) for \pb, and (0.55$\pm0.04 \mu_B$) for \li.
As the ordered moment in the ground state decreases for this series of 
compounds, the integrated intensity under the diffuse component was found to 
increase. 

Exchange constants are usually determined via the excitations from the ordered phase.
However, the large single-crystal samples required for inelastic neutron scattering
were not available for any of these compounds.
The novelty in our approach was to model $S(Q)$ in polycrystalline materials
by using HTSE of the static susceptibility.
In the determination of the NN and NNN exchange constants we used the 1$^{st}$ order
HTSE given by:
\begin{equation}
   \label{equ:htse}
    S(\vect{Q})\sim S(S+1) \left[1+\frac{S(S+1)}{3}\cdot\frac{J({\vect{Q}})}{\mathrm{k}T}\right]\cdot F^2(Q)
 \end{equation}
where $S$ is the spin of the magnetic ions, $F^2(Q)$ the magnetic form factor
of the individual magnetic ions and $J(\vect{Q})$ is the Fourier transform of the exchange interactions \cite{ashcroftmermin}.
Fig.~\ref{fig:d7}b shows the diffuse neutron scattering intensity at $T=20~\mathrm{K}$,
a temperature well into the paramagnetic phase for this compound.
Although magnetic long range order has been destroyed at this temperature,
the k$T$ term has not completely washed out all magnetic interactions.
The presence of oscillations in the data indicates the existence of short-range spin
correlations.
Note that we have measured and fixed $F^2(Q)$ directly at room temperature, where no oscillations are present and these systems are ideal paramagnets.
The calculated scattering from the two pairs of $J$'s in Fig.~\ref{fig:phase_diag} are shown in
Fig.~\ref{fig:d7}b. Our results again select the unique solution in the CAF region, with FM $J_1=(-3.2\pm1)~\mathrm{K}$ and AF $J_2=(7.7\pm1)~\mathrm{K}$.
Similarly, for \sr, the combined neutron and suscebtibility measurements yield
$J_1=(-4.6\pm1)~\mathrm{K}$ (FM) and $J_2=(6.1\pm1)~\mathrm{K}$ (AF).
Note that a similar analysis for \li~was not possible due to absorption of neutrons by the sample.

The ordered moments in the ground states are smaller than the nominal value 
for $S=1/2$ systems due to quantum fluctuations. The value for \li~is consistent with Ref. \cite{bombardi_prl04}.
However, the values for \pb~and \sr~are significantly lower, indicating an 
increase in quantum disorder.
This is entirely consistent with the fact that values of \jone~and \jtwo~determined in this study place them successively closer to the disordered region of the phase diagram.

In summary, we have unambiguously determined the exchange constants that are 
key to understanding a new region of the \jj~phase diagram with FM NN 
interactions and AF NNN interactions. We are able to tie down the exchange constants using novel polarized 
neutron scattering techniques from powdered samples. The reduced ordered moment and increased
diffuse magnetic signal indicate the approach to a quantum disordered spin-nematic 
region. It should be possible to 
tune the exchange to be closer to the quantum disordered region by other 
chemical substitutions, or by the application of pressure or magnetic field.
These low energy scale systems are particularly well suited for further experimental
study, for example in high magnetic fields \cite{schmidt_jpcm07,schmidt_prb07}, 
and the polarized neutron scattering techniques described here open new 
vistas for the exploration of these model magnetic systems.

\bibliographystyle{apsrev}

\begin{thebibliography}{31}
\expandafter\ifx\csname natexlab\endcsname\relax\def\natexlab#1{#1}\fi
\expandafter\ifx\csname bibnamefont\endcsname\relax
  \def\bibnamefont#1{#1}\fi
\expandafter\ifx\csname bibfnamefont\endcsname\relax
  \def\bibfnamefont#1{#1}\fi
\expandafter\ifx\csname citenamefont\endcsname\relax
  \def\citenamefont#1{#1}\fi
\expandafter\ifx\csname url\endcsname\relax
  \def\url#1{\texttt{#1}}\fi
\expandafter\ifx\csname urlprefix\endcsname\relax\def\urlprefix{URL }\fi
\providecommand{\bibinfo}[2]{#2}
\providecommand{\eprint}[2][]{\url{#2}}

\bibitem[{\citenamefont{Misguich and Lhuillier}(World Scientific, Singapore,
  2004)}]{misguich_ws04}
\bibinfo{author}{\bibfnamefont{G.}~\bibnamefont{Misguich}} \bibnamefont{and}
  \bibinfo{author}{\bibfnamefont{C.}~\bibnamefont{Lhuillier}},
  \emph{\bibinfo{title}{Frustrated Spin Systems}} (\bibinfo{publisher}{Edited
  by H.T. Diep}, \bibinfo{year}{World Scientific, Singapore, 2004}).

\bibitem[{\citenamefont{Kamihara et~al.}(2008)}]{kamihara_jacs08}
\bibinfo{author}{\bibfnamefont{Y.}~\bibnamefont{Kamihara}}
  \bibnamefont{et~al.}, \bibinfo{journal}{\rm{\it{J. Am. Chem. Soc. }}}
  \textbf{\bibinfo{volume}{130}}, \bibinfo{pages}{3296} (\bibinfo{year}{2008}).

\bibitem[{\citenamefont{Chen et~al.}(2008)}]{chen_nat08}
\bibinfo{author}{\bibfnamefont{X.~H.} \bibnamefont{Chen}} \bibnamefont{et~al.},
  \bibinfo{journal}{\rm{\it{Nature }}} \textbf{\bibinfo{volume}{453}},
  \bibinfo{pages}{761} (\bibinfo{year}{2008}).

\bibitem[{\citenamefont{Si and Abrahams}(2008)}]{si_prl08}
\bibinfo{author}{\bibfnamefont{Q.}~\bibnamefont{Si}} \bibnamefont{and}
  \bibinfo{author}{\bibfnamefont{E.}~\bibnamefont{Abrahams}},
  \bibinfo{journal}{\rm{\it{Phys. Rev. Lett. }}}
  \textbf{\bibinfo{volume}{101}}, \bibinfo{pages}{076401}
  (\bibinfo{year}{2008}).

\bibitem[{\citenamefont{Schulz et~al.}(1996)\citenamefont{Schulz, Ziman, and
  Poilblanc}}]{schulz_jpif96}
\bibinfo{author}{\bibfnamefont{H.~J.} \bibnamefont{Schulz}},
  \bibinfo{author}{\bibfnamefont{T.~A.~L.} \bibnamefont{Ziman}},
  \bibnamefont{and}
  \bibinfo{author}{\bibfnamefont{D.}~\bibnamefont{Poilblanc}},
  \bibinfo{journal}{\rm{\it{J. Phys. I France}}} \textbf{\bibinfo{volume}{6}},
  \bibinfo{pages}{675} (\bibinfo{year}{1996}).

\bibitem[{\citenamefont{Shannon et~al.}(2004)}]{shannon_epjb04}
\bibinfo{author}{\bibfnamefont{N.}~\bibnamefont{Shannon}} \bibnamefont{et~al.},
  \bibinfo{journal}{\rm{\it{Eur. Phys. J. B }}} \textbf{\bibinfo{volume}{41}},
  \bibinfo{pages}{599} (\bibinfo{year}{2004}).

\bibitem[{\citenamefont{Shannon et~al.}(2006)\citenamefont{Shannon, Momoi, and
  Sindzingre}}]{shannon_prl06}
\bibinfo{author}{\bibfnamefont{N.}~\bibnamefont{Shannon}},
  \bibinfo{author}{\bibfnamefont{T.}~\bibnamefont{Momoi}}, \bibnamefont{and}
  \bibinfo{author}{\bibfnamefont{P.}~\bibnamefont{Sindzingre}},
  \bibinfo{journal}{\rm{\it{Phys. Rev. Lett. }}} \textbf{\bibinfo{volume}{96}},
  \bibinfo{pages}{027213} (\bibinfo{year}{2006}).

\bibitem[{\citenamefont{Millet and Satto}(1998)}]{millet_mrb98}
\bibinfo{author}{\bibfnamefont{P.}~\bibnamefont{Millet}} \bibnamefont{and}
  \bibinfo{author}{\bibfnamefont{C.}~\bibnamefont{Satto}},
  \bibinfo{journal}{\rm{\it{Materials Research Bulletin }}}
  \textbf{\bibinfo{volume}{33}}, \bibinfo{pages}{1339} (\bibinfo{year}{1998}).

\bibitem[{\citenamefont{Melzi et~al.}(2000)}]{melzi_prl00}
\bibinfo{author}{\bibfnamefont{R.}~\bibnamefont{Melzi}} \bibnamefont{et~al.},
  \bibinfo{journal}{\rm{\it{Phys. Rev. Lett. }}} \textbf{\bibinfo{volume}{85}},
  \bibinfo{pages}{1318} (\bibinfo{year}{2000}).

\bibitem[{\citenamefont{Melzi et~al.}(2001)}]{melzi_prb01}
\bibinfo{author}{\bibfnamefont{R.}~\bibnamefont{Melzi}} \bibnamefont{et~al.},
  \bibinfo{journal}{\rm{\it{Phys. Rev. B }}} \textbf{\bibinfo{volume}{64}},
  \bibinfo{pages}{024409} (\bibinfo{year}{2001}).

\bibitem[{\citenamefont{Carretta
  et~al.}(2002{\natexlab{a}})\citenamefont{Carretta, Melzi, Papinutto, and
  Millet}}]{carretta_prl02}
\bibinfo{author}{\bibfnamefont{P.}~\bibnamefont{Carretta}},
  \bibinfo{author}{\bibfnamefont{R.}~\bibnamefont{Melzi}},
  \bibinfo{author}{\bibfnamefont{N.}~\bibnamefont{Papinutto}},
  \bibnamefont{and} \bibinfo{author}{\bibfnamefont{P.}~\bibnamefont{Millet}},
  \bibinfo{journal}{\rm{\it{Phys. Rev. Lett. }}} \textbf{\bibinfo{volume}{88}},
  \bibinfo{pages}{047601} (\bibinfo{year}{2002}{\natexlab{a}}).

\bibitem[{\citenamefont{Carretta et~al.}(2002{\natexlab{b}})}]{carretta_prb02}
\bibinfo{author}{\bibfnamefont{P.}~\bibnamefont{Carretta}}
  \bibnamefont{et~al.}, \bibinfo{journal}{\rm{\it{Phys. Rev. B }}}
  \textbf{\bibinfo{volume}{66}}, \bibinfo{pages}{094420}
  (\bibinfo{year}{2002}{\natexlab{b}}).

\bibitem[{\citenamefont{Rosner et~al.}(2002)}]{rosner_prl02}
\bibinfo{author}{\bibfnamefont{H.}~\bibnamefont{Rosner}} \bibnamefont{et~al.},
  \bibinfo{journal}{\rm{\it{Phys. Rev. Lett. }}} \textbf{\bibinfo{volume}{88}},
  \bibinfo{pages}{186405} (\bibinfo{year}{2002}).

\bibitem[{\citenamefont{Rosner et~al.}(2003)\citenamefont{Rosner, Singh, Zheng,
  Oitmaa, and Pickett}}]{rosner_prb03}
\bibinfo{author}{\bibfnamefont{H.}~\bibnamefont{Rosner}},
  \bibinfo{author}{\bibfnamefont{R.~R.~P.} \bibnamefont{Singh}},
  \bibinfo{author}{\bibfnamefont{W.~H.} \bibnamefont{Zheng}},
  \bibinfo{author}{\bibfnamefont{J.}~\bibnamefont{Oitmaa}}, \bibnamefont{and}
  \bibinfo{author}{\bibfnamefont{W.~E.} \bibnamefont{Pickett}},
  \bibinfo{journal}{\rm{\it{Phys. Rev. B }}} \textbf{\bibinfo{volume}{67}},
  \bibinfo{pages}{014416} (\bibinfo{year}{2003}).

\bibitem[{\citenamefont{Bombardi et~al.}(2004)}]{bombardi_prl04}
\bibinfo{author}{\bibfnamefont{A.}~\bibnamefont{Bombardi}}
  \bibnamefont{et~al.}, \bibinfo{journal}{\rm{\it{Phys. Rev. Lett. }}}
  \textbf{\bibinfo{volume}{93}}, \bibinfo{pages}{027202}
  (\bibinfo{year}{2004}).

\bibitem[{\citenamefont{Bombardi et~al.}(2005)}]{bombardi_prb05}
\bibinfo{author}{\bibfnamefont{A.}~\bibnamefont{Bombardi}}
  \bibnamefont{et~al.}, \bibinfo{journal}{\rm{\it{Phys. Rev. B }}}
  \textbf{\bibinfo{volume}{71}}, \bibinfo{pages}{220406(R)}
  (\bibinfo{year}{2005}).

\bibitem[{\citenamefont{Tsirlin et~al.}(2008)}]{tsirlin_prb08}
\bibinfo{author}{\bibfnamefont{A.~A.} \bibnamefont{Tsirlin}}
  \bibnamefont{et~al.}, \bibinfo{journal}{\rm{\it{Phys. Rev. B }}}
  \textbf{\bibinfo{volume}{77}}, \bibinfo{pages}{092402}
  (\bibinfo{year}{2008}).

\bibitem[{\citenamefont{Oka et~al.}(2008)}]{oka_ic08}
\bibinfo{author}{\bibfnamefont{K.}~\bibnamefont{Oka}} \bibnamefont{et~al.},
  \bibinfo{journal}{\rm{\it{Inorg. Chem. }}} \textbf{\bibinfo{volume}{47}},
  \bibinfo{pages}{7355} (\bibinfo{year}{2008}).

\bibitem[{\citenamefont{Kaul et~al.}(2004)}]{kaul_jmmm04}
\bibinfo{author}{\bibfnamefont{E.~E.} \bibnamefont{Kaul}} \bibnamefont{et~al.},
  \bibinfo{journal}{\rm{\it{J. Magn. Magn. Mater. }}}
  \textbf{\bibinfo{volume}{272-276}}, \bibinfo{pages}{922}
  (\bibinfo{year}{2004}).

\bibitem[{\citenamefont{Shpanchenko et~al.}(2006)}]{shpanchenko_acsc06}
\bibinfo{author}{\bibfnamefont{R.~V.} \bibnamefont{Shpanchenko}}
  \bibnamefont{et~al.}, \bibinfo{journal}{\rm{\it{Acta Cryst. C, }}}
  \textbf{\bibinfo{volume}{62}}, \bibinfo{pages}{i88} (\bibinfo{year}{2006}).

\bibitem[{\citenamefont{Skoulatos et~al.}(2007)}]{skoulatos_jmmm07}
\bibinfo{author}{\bibfnamefont{M.}~\bibnamefont{Skoulatos}}
  \bibnamefont{et~al.}, \bibinfo{journal}{\rm{\it{J. Magn. Magn. Mater. }}}
  \textbf{\bibinfo{volume}{310}}, \bibinfo{pages}{1257} (\bibinfo{year}{2007}).

\bibitem[{\citenamefont{Nath et~al.}(2008)\citenamefont{Nath, Tsirlin, Rosner,
  and Geibel}}]{nath_prb08}
\bibinfo{author}{\bibfnamefont{R.}~\bibnamefont{Nath}},
  \bibinfo{author}{\bibfnamefont{A.~A.} \bibnamefont{Tsirlin}},
  \bibinfo{author}{\bibfnamefont{H.}~\bibnamefont{Rosner}}, \bibnamefont{and}
  \bibinfo{author}{\bibfnamefont{C.}~\bibnamefont{Geibel}},
  \bibinfo{journal}{\rm{\it{Phys. Rev. B }}} \textbf{\bibinfo{volume}{78}},
  \bibinfo{pages}{064422} (\bibinfo{year}{2008}).

\bibitem[{\citenamefont{Kageyama et~al.}(2005)}]{kageyama_jpsj05}
\bibinfo{author}{\bibfnamefont{H.}~\bibnamefont{Kageyama}}
  \bibnamefont{et~al.}, \bibinfo{journal}{\rm{\it{J. Phys. Soc. Jpn. }}}
  \textbf{\bibinfo{volume}{74}}, \bibinfo{pages}{1702} (\bibinfo{year}{2005}).

\bibitem[{\citenamefont{Oba et~al.}(2006)}]{oba_jpsj06}
\bibinfo{author}{\bibfnamefont{N.}~\bibnamefont{Oba}} \bibnamefont{et~al.},
  \bibinfo{journal}{\rm{\it{J. Phys. Soc. Jpn. }}}
  \textbf{\bibinfo{volume}{75}}, \bibinfo{pages}{113601}
  (\bibinfo{year}{2006}).

\bibitem[{\citenamefont{Thalmeier et~al.}(2008)\citenamefont{Thalmeier,
  Zhitomirsky, Schmidt, and Shannon}}]{thalmeier_prb08}
\bibinfo{author}{\bibfnamefont{P.}~\bibnamefont{Thalmeier}},
  \bibinfo{author}{\bibfnamefont{M.~E.} \bibnamefont{Zhitomirsky}},
  \bibinfo{author}{\bibfnamefont{B.}~\bibnamefont{Schmidt}}, \bibnamefont{and}
  \bibinfo{author}{\bibfnamefont{N.}~\bibnamefont{Shannon}},
  \bibinfo{journal}{\rm{\it{Phys. Rev. B }}} \textbf{\bibinfo{volume}{77}},
  \bibinfo{pages}{104441} (\bibinfo{year}{2008}).

\bibitem[{\citenamefont{{O. Sch\"{a}rpf and H.
  Capellmann}}(1993)}]{scharpf_pss93}
\bibinfo{author}{\bibnamefont{{O. Sch\"{a}rpf and H. Capellmann}}},
  \bibinfo{journal}{\rm{\it{Phys. Stat. Sol. }}}
  \textbf{\bibinfo{volume}{A135}}, \bibinfo{pages}{359} (\bibinfo{year}{1993}).

\bibitem[{\citenamefont{Kaul}(2005)}]{kaul_thesis}
\bibinfo{author}{\bibfnamefont{E.~E.} \bibnamefont{Kaul}},
  \emph{\bibinfo{title}{``Experimental Investigation of New Low-Dimensional
  Spin Systems in Vanadium Oxides''}} (\bibinfo{publisher}{PhD
  Thesis,~Technische Universit\"{a}t Dresden, Dresden}, \bibinfo{year}{2005}).

\bibitem[{\citenamefont{{S. Meyer and B. Mertens and Hk.
  M\"{u}ller-Buschbaum}}(1997)}]{meyer_b}
\bibinfo{author}{\bibnamefont{{S. Meyer and B. Mertens and Hk.
  M\"{u}ller-Buschbaum}}}, \bibinfo{journal}{\rm{\it{Zeitschrift f\uumlaut r
  Naturforschung }}} \textbf{\bibinfo{volume}{52b}}, \bibinfo{pages}{985}
  (\bibinfo{year}{1997}).

\bibitem[{\citenamefont{Ashcroft and Mermin}(1976)}]{ashcroftmermin}
\bibinfo{author}{\bibfnamefont{N.~W.} \bibnamefont{Ashcroft}} \bibnamefont{and}
  \bibinfo{author}{\bibfnamefont{N.~D.} \bibnamefont{Mermin}},
  \emph{\bibinfo{title}{Solid State Physics, ch. 33}}
  (\bibinfo{publisher}{Holt-Saunders International Editions},
  \bibinfo{year}{1976}).

\bibitem[{\citenamefont{Schmidt
  et~al.}(2007{\natexlab{a}})\citenamefont{Schmidt, Shannon, and
  Thalmeier}}]{schmidt_jpcm07}
\bibinfo{author}{\bibfnamefont{B.}~\bibnamefont{Schmidt}},
  \bibinfo{author}{\bibfnamefont{N.}~\bibnamefont{Shannon}}, \bibnamefont{and}
  \bibinfo{author}{\bibfnamefont{P.}~\bibnamefont{Thalmeier}},
  \bibinfo{journal}{\rm{\it{J. Phys.: Condens. Matter }}}
  \textbf{\bibinfo{volume}{19}}, \bibinfo{pages}{145211}
  (\bibinfo{year}{2007}{\natexlab{a}}).

\bibitem[{\citenamefont{Schmidt
  et~al.}(2007{\natexlab{b}})\citenamefont{Schmidt, Thalmeier, and
  Shannon}}]{schmidt_prb07}
\bibinfo{author}{\bibfnamefont{B.}~\bibnamefont{Schmidt}},
  \bibinfo{author}{\bibfnamefont{P.}~\bibnamefont{Thalmeier}},
  \bibnamefont{and} \bibinfo{author}{\bibfnamefont{N.}~\bibnamefont{Shannon}},
  \bibinfo{journal}{\rm{\it{Phys. Rev. B }}} \textbf{\bibinfo{volume}{76}},
  \bibinfo{pages}{125113} (\bibinfo{year}{2007}{\natexlab{b}}).

\end{thebibliography}
\end{document}